\documentstyle[11pt,newpasp,twoside,epsf]{article}
\def\edcomment#1{\iffalse\marginpar{\raggedright\sl#1\/}\else\relax\fi}
\reversemarginpar

\begin{document}
\title{Far-UV FUSE Spectra of Peculiar Magnetic Cataclysmic
 Variables}
 \author{Martine Mouchet$^1$, Jean-Marc Bonnet-Bidaud$^2$, Evelyne Roueff$^1$, 
Meil Abada-Simon$^1$, Klaus Beuermann$^3$, Domitilla de Martino$^4$, 
Jean-Michel Desert$^5$,  Roger Ferlet$^5$, Robert Fried$^6$, Boris 
G\"ansicke$^7$, Steve Howell$^8$, Koji Mukai$^9$, Delphine Porquet$^{10}$,
Paula Szkody$^{11}$\\}

\bigskip
\affil{$^1$Observatoire de Paris, 5 Place J. Janssen, 92190 Meudon, F\\
$^2$DAPNIA/Service d'Astrophysique,CEA-Saclay, Gif-sur-Yvette, F\\
$^3$Universt\"ats-Sternwarte, G\"ottingen, D \\
$^4$Osservatorio Astronomico  Capodimonte, Naples, I\\
$^5$Institut d'Astrophysique de Paris, Paris, F\\
$^6$Braeside Observatory, Flagstaff, AZ, USA\\
$^7$Department of Physics and Astronomy, University of Southampton, UK\\
$^8$Astrophysics Group, Planetary Science Institute, Tucson, AZ, USA\\
$^9$NASA Goddard Space Flight Center,  Greenbelt, MD,  USA\\
$^{10}$Max-Planck-Institut für Extraterrestrische Physik, Garching, D\\
$^{11}$Department of Astronomy, University of Washington, Seattle, WA, USA\\}
\bigskip

\begin{abstract}
We present far-UV spectra of the three magnetic cataclysmic 
variables (MCVs) BY Cam, V1309 Ori and AE Aqr obtained with 
the FUSE satellite. These MCVs have revealed strongly unusual 
NV and CIV UV resonance lines. The FUSE spectra  exhibit broad 
OVI lines as well as a strong NIII line at 991\AA, while the 
CIII 1175\AA{ } line is nearly absent, supporting non-solar CNO 
abundances of the accreting matter in these sources. The spectrum of BY Cam
shows molecular H$_2$ lines which might be of circumstellar nature. The flaring
activity of AE Aqr is also observed in the far-UV range. The radial velocities
of the broad OVI components in AE Aqr are orbitally modulated and 
would indicate an emission region close to the magnetosphere.
\end{abstract}

\section{Introduction}
Among magnetic cataclysmic variables (MCV), three Polars (BY Cam, 
V1309 Ori, MN Hya)
 and one intermediate polar (AE Aqr)  have been discovered to exhibit 
very peculiar NV and CIV UV resonance line intensities, with a NV/CIV 
ratio much greater
than the typical value\,$\sim$\,0.5 (Bonnet-Bidaud \& Mouchet 1987 (BM87),
Skzody \& Silber 1996, Schmidt \& Stockman 2001, Jameson et al. 1980).
Anomalous UV line strengths might result from  non-solar abundances in the 
accreting matter or from peculiar ionization conditions
in the emitting regions (BM87, Mouchet et al. 2003).
In order to have access to the OVI resonance lines at 1032-1038\AA, 
we obtained FUSE spectra of three of these sources, BY Cam, 
V1309 Ori and AE Aqr. 

In addition to these abnormal UV line intensities, all three sources
exhibit peculiarities. 
BY Cam is enigmatic in several respects, being slightly  
de-synchronized, and possessing a 
very high hard X-ray-to-optical luminosity ratio (Kallman et al. 1996).
Among more than fifty Polars, only five, including the nova magnetic system 
V1500 Cyg, have been found
to be slightly de-synchronized.   BY Cam has two close  periods near
3.3\,h that differ by  only $\sim$ 1\%  
(Silber et al. 1992, Mouchet et al. 1997).
 V1309 Ori is also very peculiar, having the longest 
orbital period for a Polar
(P$_{\rm orb}=7.98$\,h, Garnavich et al. 1994, Walter et al. 1995), 
revealing a very high soft-to-hard X-ray luminosity ratio and a strong 
flaring X-ray variability (Walter et al. 1995, de Martino et al. 1998). 
AE Aqr exhibits several unique properties (see e.g. detailed 
introduction in Welsh et al. 1998).  It has  been detected in all
energy bands, from TeV to radio wavelengths. This binary displays a wide 
range of variability, both coherent and incoherent, on various timescales. 
The spin period of the white dwarf (33\,s)  is the fastest  among IPs
 and the spin-to-orbital period ratio (33s/9.88h $\sim$ 10$^{-3}$) is
the smallest.  Strong aperiodic recurrent flares of about one-hour 
duration have been recorded in all wavelength bands. To account for this 
strong variability and a very low X-ray luminosity,  Wynn et al. (1997)
proposed that the gas flow consists of blobs which are propelled out of the
system by the rapidly rotating magnetosphere. 

\section{Average FUSE spectra}
\begin{figure}
\plotfiddle{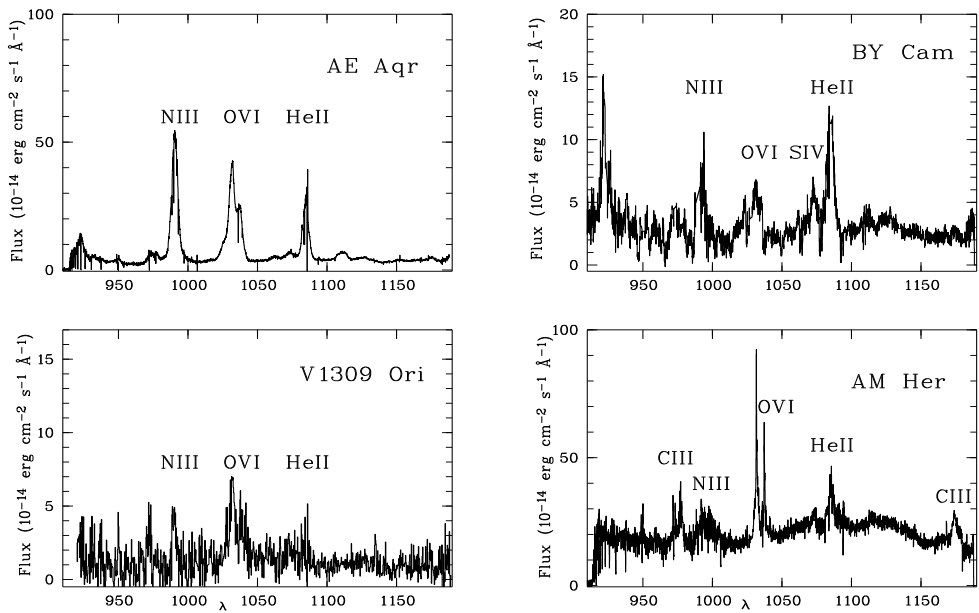}{8.cm}{90}{50}{45}{210}{-20}
\caption{Average FUSE spectra for AE Aqr (upper left), BY Cam (upper right),
V1309 Ori (lower left) and AM Her for comparison (lower right). }
\end{figure}

Far UV observations at high spectral resolution in the range 905-1185\AA{ } 
have been performed with the FUSE satellite  using the time-tag
 mode and the large aperture ($30\arcsec \times 30\arcsec$) 
(Moos et al. 2000). Total 
exposure times of 20\,ks, 12\,ks and 40\,ks split into 5, 5 and 14 exposures 
were obtained respectively for  BY Cam, V1309 Ori and AE Aqr.
The spectra of BY Cam mostly acquired during ``day time'' are strongly 
affected by geocoronal lines.

The average spectra obtained by combining all detectors and exposures
are reported in Fig.1 where the strongest  geocoronal lines have been
removed.  For comparison the FUSE average spectrum of AM Her 
(Hutchings et al. 2002) retrieved from the archives, is also plotted.
The level of the continuum is consistent 
with the extrapolation of the power law fitting the IUE or HST 
spectra obtained during high states of accretion for the two polars,
and during quiescent states for AE Aqr.
 Broad emission lines are present, including
the resonance doublet OVI and strong lines of  HeII and NIII.
Note the absence of the CIII 1175\AA{ } line and the weakness of CIII
at 975\AA{}. 
Interstellar lines of ArI, FeII and SiII are detected in BY Cam, and of 
HI, CII, OI and NII in AE Aqr.
Pronounced H$_2$ molecular lines are also clearly 
present in the spectrum of BY Cam (see Section 3).
Contrary to AM Her, a narrow component in the OVI line is not seen in
BY Cam, and is much weaker in AE Aqr, while the spectrum of V1309 Ori is 
too noisy to distinguish several components in the OVI profile.  
The dereddened total OVI line intensities are respectively of 
2.5$\times$10$^{-13}$, 4$\times$10$^{-13}$
and 27$\times$10$^{-13}$\,erg\,s$^{-1}$\,cm$^{-2}$ for BY Cam, V1309 Ori and 
AE Aqr. In the case of BY Cam, 
the redward component of the doublet is strongly affected by the presence 
of H$_2$ molecular lines. 
We have derived the flux of this line taking into account
this strong absorption (Mouchet et al. 2003).
A modelisation of the resonance UV CNO lines measured in BY Cam 
is presented in Bonnet-Bidaud \& Mouchet (these proceedings).

\section{ H$_2$ molecular lines in BY Cam }
\begin{figure}
\plotfiddle{ 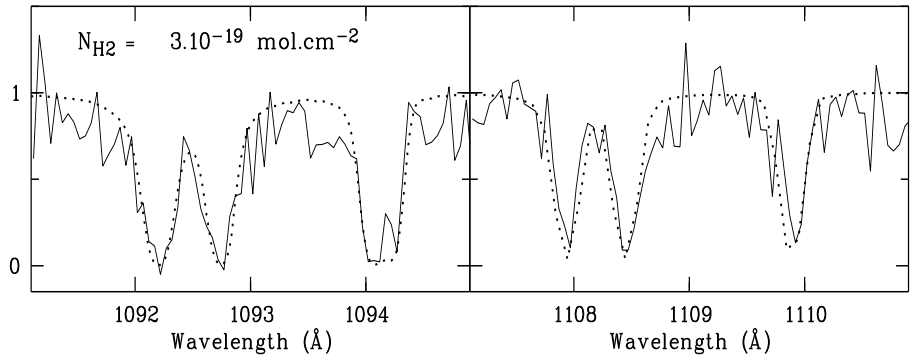}{5.5cm}{-90}{50}{60}{-220}{+350}
\caption{H$_2$ absorption lines corresponding to the 0-0
and 1-0 vibration transitions  as observed in the normalized spectrum 
of BY Cam (full lines) and as
derived from the best fit solution with  N$_{\rm H_2}$ = 3$\times 10^{19}$ 
molecules\,cm$^{-2}$ (dotted lines). }
\end{figure}

Several H$_2$ absorption lines of the Lyman and Werner 
bands are clearly detected in the average FUSE spectrum of BY Cam. 
The equivalent width measurements of the strongest  H$_2$  lines 
absorbed from the J=0, 1, and 2 levels in the v'-0 Lyman bands 
with v'\,=\,0 to 4, have been compared with those of  models computed
for specific values of the
total column density, a temperature of 80\,K and a Doppler factor of 
10\,km\,s$^{-1}$, using recent molecular data given by Abgrall et al. (2000).
A best fit value is obtained 
for a H$_2$ column density N$_{\rm H_2}$ of  $3\times 10^{19}$ 
molecules\,cm$^{-2}$. 
This  model for the first lines of the Lyman (0,0) and (1,0) 
bands is displayed in Fig.2, together with 
the corresponding parts of the observed spectrum.
From the absorption bump around 2200 \AA, an upper limit of 
E$_{\rm B-V}\leq$ 0.05 has been derived for the source (BM87), which 
corresponds to a neutral H density 
N$_{\rm H_1} ~\leq ~2.6\times 10^{20}$ cm$^{-2}$ (Shull \& Van Steenberg 1985).
This value is consistent with the equivalent neutral H column density
N$_{\rm H_1} \sim (1-2)\times 10^{20}$ cm$^{-2}$ derived from
the absorption cut-off in the X-ray spectra (Kallman et al. 1996).
This implies a very high N$_{\rm H_2}$/N$_{\rm H_1}$ ratio in the 
direction of the source, which is unexpected for a relatively nearby source. 
It may be due to the presence of a molecular cloud in 
the line of sight 
but it may also be linked with dense matter close to the source, possibly 
ejected during a prior nova event. Note that such  H$_2$ absorption 
has also being detected in the FUSE spectra of the supersoft binary
QR And and may originate in a circumbinary location (Hutchings et al. 
2001). 

\begin{figure}
\plotfiddle{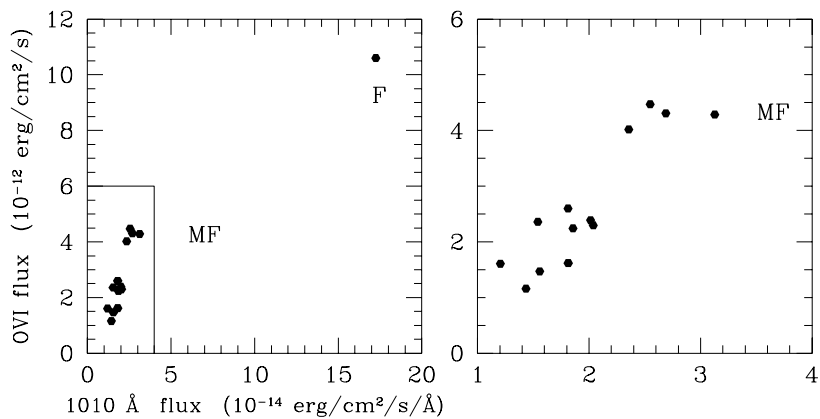}{6.cm}{0}{60}{60}{-180}{-100}
\caption{The OVI line flux versus the continuum flux at 1010\AA{ } measured in 
the forteen  AE Aqr spectra. The 
lower region is zoomed in the  right panel. Note the 
flare (F) at the upper corner of the left panel and the four 
moderate flares (MF) more visible in the upper right of the zoom.}
\end{figure}

\section{Far-UV variability of AE Aqr}

We obtained 14 spectra of AE Aqr with exposure times in the range 2500\,s
($\Delta \phi_{\rm orb}= 0.7$) to 3400\,s ($\Delta \phi_{\rm orb}= 0.10$). 
They are regularly spaced, covering 2.2 orbital cycles. In order to describe
the OVI doublet,  the profile has been fitted using one narrow 
($\sim$\,450\,km\,s$^{-1}$)  
and one broad ($\sim$\,1300 km\,s$^{-1}$) gaussian for each component.
Such components could not be detected in the HST low resolution spectra 
(Eracleous \& Horne 1996). 
Another broad component has also been added
to take into account the Ly$\beta$ contribution as well as  an absorption 
contribution for the interstellar CII line at 1036\AA. 
All phases quoted below refer to Casares et al. (1996) ephemeris with zero 
phase defined as the time at which the secondary is closest to the 
observer (blue-to-red crossing of  the secondary absorption line 
radial velocities).

\begin{figure}
\plotfiddle{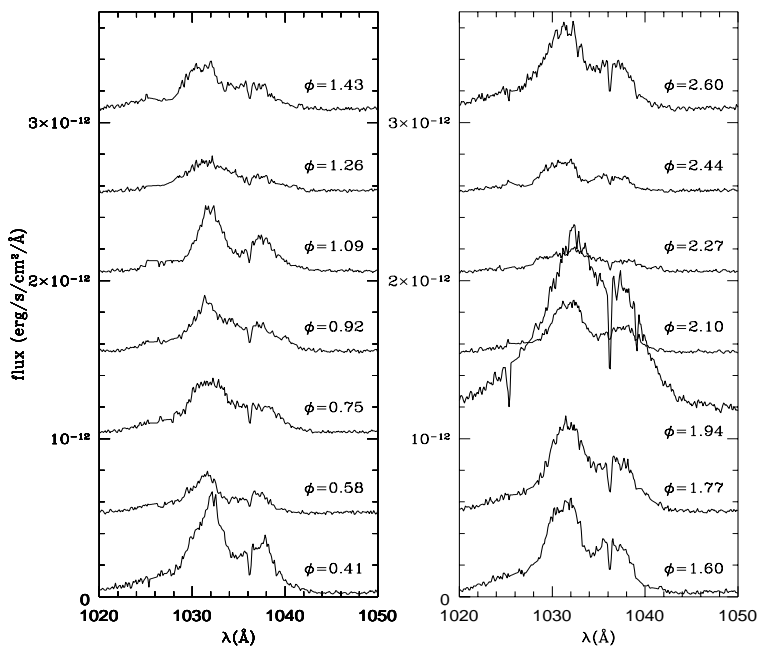}{8.5cm}{0}{50}{45}{-170}{-70}
\caption{The OVI doublet for the 14 FUSE spectra of AE Aqr. The spectra are
 offset by 5.1$\times$10$^{-13}$ erg\,s$^{-1}$cm$^{-2}$\,\AA$^{-1}$. 
Orbital phases are indicated. Note the strong flare at phase 1.94.}
\end{figure}

\subsection{Flaring activity }

 In Fig.\,3 the continuum measured at 1010\AA{}\, is plotted versus the 
OVI line 
intensity and the OVI line profiles of the 14 spectra
are displayed in Fig.\,4.  A large flare occurs  at 
$\phi$=1.94, with an increase by a factor 9 in the 
continuum flux and 5 in
the OVI line flux. The line width is also 
enhanced up to 2400 km\,s$^{-1}$. 
Both broad and narrow components are more intense and wider.
A possible delay between the lines and the 
continuum, as observed in the optical, but not in the HST data, will be 
investigated by reconstructing spectra at higher temporal 
resolution.
In addition the continuum and line fluxes increase by a factor 1.5-2 
during four other exposures ($\phi$\,=\,0.41, 1.60, 1.77, 2.60).
These moderate flares are quite similar to those observed in the UV range
with IUE (de Martino et al. 1995).

\subsection{Search for orbital variability}
The  continuum flux of the remaining nine quiescent spectra  shows a 
weak tendency for a minimum at phase 0.0, similar to the optical light
curve. The narrow component intensity also tends to be lower at this 
phase, while the broad component flux does not show any regular 
orbital modulation. 
The radial velocities (RVs) have been derived for all components used to 
describe the OVI profile.
While the interstellar CII RVs are stable 
(dispersion less than 5 km\,s$^{-1}$), those of the OVI broad components 
are modulated.
The corresponding sinusoidal parameters
are  K\,=\,192$\,\pm\, 30$ km\,s$^{-1}$, $\gamma\,=\,-93\,\pm\, 8$ km\,s$^{-1}$
and $\phi_0\,=\,0.74\,\pm\,0.05$ where   $\phi_0$ is the blue-to-red 
crossing phase. Note that the RV of the flare spectrum  does not depart from
the orbital modulation.
These values are marginally consistent with those obtained for the UV 
lines (K $\sim$ 110-330  km\,s$^{-1}$, $\phi_0\, \sim$\,0.62-0.73,
 Eracleous \& Horne 1996). However they  contrast with the X-ray line RVs
which exhibit a much larger amplitude (see Osborne \& Wynn these proc.). \\
The RV measurements of the narrow component show much scatter with a    
$\phi_0\,$ value close to 0, indicating an emitting region  in phase with 
the secondary motion, such as the irradiated hemisphere of the companion.
The exact location cannot be ascertained without a precise evaluation of the 
K and $\gamma$ velocities. \\

While the phasing and the RV amplitude of the broad OVI components  are 
consistent with
blobby gas having passed the closest approach to the white dwarf
(Wynn et al. 1997), the large width of the
OVI line would require a significant contribution of 
individual blobs at high velocities. Such blobs are not predicted  
in large numbers by the propeller-driven collisional model (Welsh, Horne
\& Gomer 1998). Besides such high velocities are easily reached close to 
the magnetosphere radius.  
In addition, the UV line intensities can be accounted for by material 
collisionally heated 
at temperatures $\sim$ 5\,10$^5$\,K, however with a non-solar chemical 
composition (Mouchet et al. in prep.). Such abundances can be obtained as
the result of nuclear evolution of the donor in the framework of evolutionary
models in which AE Aqr would descend from a 
supersoft X-ray binary (Schenker et al. 2002, and these proc.)  \\
\medskip

\acknowledgments

We are very grateful to Sandrine Maloreau for her participation in the 
reduction and analysis of the AE Aqr spectra.

\end{document}